\documentclass[aps,pra,twocolumn,groupedaddress]{revtex4}
\usepackage{graphicx}
\usepackage{setspace}
\usepackage{amsmath}
\usepackage{amssymb}
\usepackage{color}
\usepackage{array}
\usepackage{subfigure}
\usepackage{hyperref}
\usepackage{float}
\begin{document}
\title{The stochastic nature induced by laser noise in narrow transitions}
\author{Yuan Sun}
\email{yuan.sun.1@stonybrook.edu}
\affiliation{Department of Physics and Astronomy, Stony Brook University, Stony Brook, NY 11794-3800, USA}
\author{Chen Zhang}
\affiliation{Department of Physics and JILA, University of Colorado, Boulder, Colorado 80309-0440, USA}
\affiliation{Department of Physics, Purdue University, West Lafayette, Indiana 47906, USA}
\date{\today}
\begin{abstract}
We use a probability--theory approach to study the laser noise's effects on laser--atom interactions. We consider the case where the atom is described by a two--level system without spontaneous emission and the laser has both intensity and frequency noises. A stochastic differential equation is established based on the Schr\"odinger equation of the laser--atom interaction in the semiclassical picture. We then analyze the equation using the path--integral technique to the first order of a perturbation approach. Because of the presence of laser noises, the atom wave function at a given time is a random variable. Therefore we construct a stochastic process charactering its time evolution. We also provide the theoretical description for the experimental realization of measuring the laser linewidth by driving a narrow atomic transition.
\end{abstract}
\maketitle

The time evolution of a two--level atom interacting with a monochromatic radiation field is a standard part of current atomic physics textbooks. However,  monochromatic electromagnetic radiation is merely an idealization of experimental realities. A laser has both intensity and frequency fluctuations that make it spectrally broaden.
A natural starting point for incorporating the effect of laser noises into the description of the laser--atom interaction is the use of the stochastic differential equation (SDE). The typical method to deal with this kind of SDE is the quantum master equation \cite{PhysRevA.31.3761}, which also finds applications in many other fields in treating noise and robustness related problems \cite{PhysRevA.77.052305}. This topic has been extensively explored in the 1970s and 1980s, both experimentally \cite{PhysRevLett.53.439, PhysRevA.36.178, PhysRevA.41.2580, PhysRevLett.64.1346} and theoretically\cite{PhysRevA.38.4657, PhysRevA.21.1289, PhysRevA.18.1490, PhysRevLett.37.1387, 0022-3700-10-2-006, PhysRevA.19.1151, PhysRevA.17.1547, PhysRevLett.42.1609}. The representation of laser noises in the SDE is usually multiplicative in nature, making the SDE difficult to solve. Yet, as previous publications have ingeniously suggested, this type of SDE can be converted into a series of ordinary differential equations to solve for one--time atom--field averages \cite{PhysRevA.20.2420}. Moreover, the spontaneous emission of the atom has always been included in those discussions.
\par
In the case of highly coherent processes where the spontaneous emission of the atom does not play an important role, a direct approach to revealing the effect of laser noise is desirable. Many interesting processes belong to this category: laser force/cooling on atoms in a time scale less than the radiative lifetime \cite{PhysRevA.77.061401} and the stimulated Raman adiabatic passage (STIRAP) \cite{PhysRevA.78.053415} \cite{PhysRevA.85.023401}. The same framework will also shed light on atomic clocks \cite{PhysRevLett.109.230801} and related precision measurement fields \cite{PhysRevLett.61.1097} where a complete characterization of noise--related effects is required \cite{PhysRev.161.350}. There is also a general interest in classical noise's effects on the Berry phase \cite{PhysRevA.87.060303, PhysRevLett.91.090404}, in which scenario a magnetic field that drives the two level spin system, rather than the laser which is the topic of this work. 
\par
In this work, we use the Schr\"odinger equation for the two--level atom wave function, where the laser--atom interaction is treated in the semiclassical description. After the corresponding SDE is established, we analyze that SDE with the Feynman's path integral for discrete states and stochastic calculus. Although the SDE is hard to solve analytically, we manage to construct an approximate solution, in the sense that it is the first--order result of a systematic perturbation approach. This perturbative treatment can be extended to higher orders and three--level or four--level atomic systems. To demonstrate the applications of this method, we calculate the atom's response to the $\pi$--pulse of a noisy laser. We also show how laser noise affects the time correlation of the atom wave function. The relation between the parameters of the laser--noise models and the laser linewidth measurement experiment is then discussed.
\par
\begin{figure*}
 \centering
\begin{tabular}{lll}
\includegraphics[width=6cm,height=4cm]{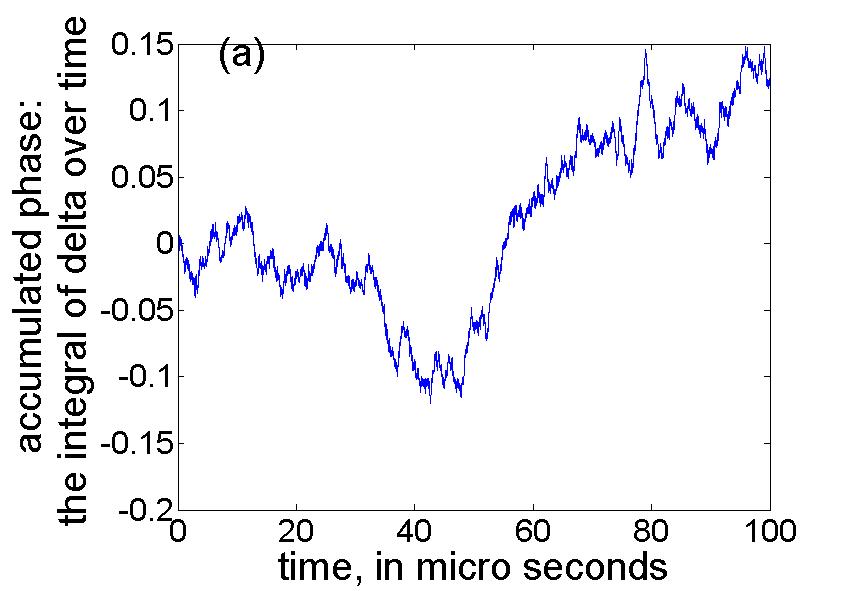} &
\includegraphics[width=6cm,height=4cm]{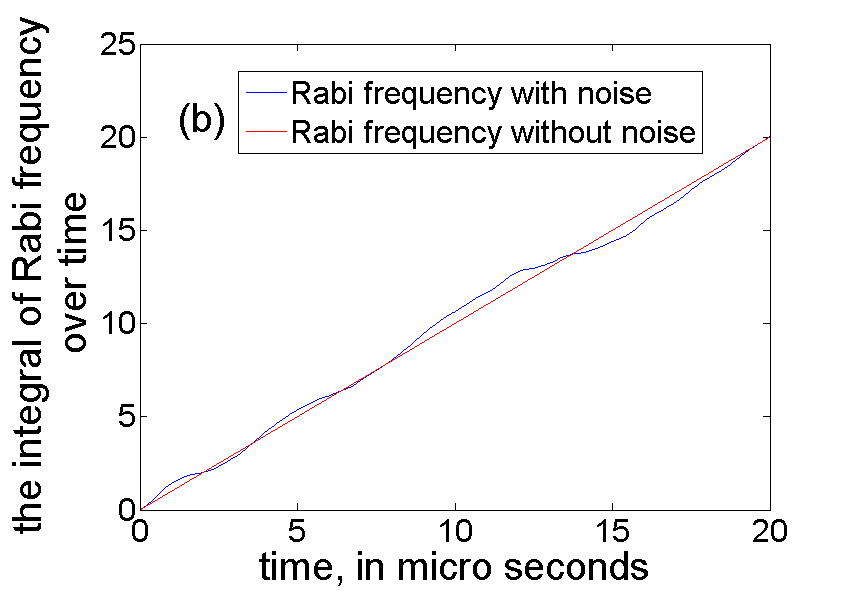} & 
\includegraphics[width=6cm,height=4cm]{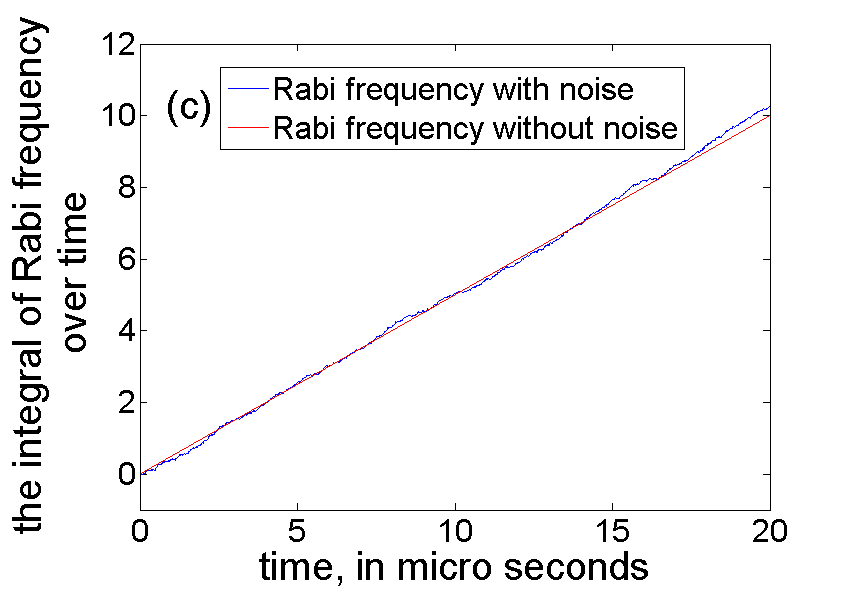}
\end{tabular}
\linespread{1} 
\caption{(Color online) Illustration for the frequency and intensity fluctuation models of a laser. (a) An example of phase diffusion: the time evolution of the phase when it is subject to a frequency fluctuation is given by the phase diffusion model. (b) The integral of Rabi frequency over time, when the intensity fluctuation obeys the stationary Ornstein-Uhlenbeck process. (c) The integral of Rabi frequency over time, given by the Brownian motion plus a constant drift. \label{noise_models}}
\end{figure*}

The classical noise of a laser can be categorized as frequency and intensity fluctuations. The frequency fluctuation is well described by the phase diffusion model \cite{PhysRevA.36.178}, while the intensity fluctuation is well described by the real Gaussian field or the complex Gaussian fields model \cite{PhysRevA.45.468}. In this work, without loss of generality, we use Brownian motion \cite{Zoller} to model the integral of the Rabi frequency over time, which simplifies the computation. Our methods work equally well when the intensity noise is described by other models, as discussed below. This simplification can also be viewed as one extreme situation of Ornstein-Uhlenbeck type noise.  Qualitative examples of the laser noise's behavior according to the above models are shown in Fig. \ref{noise_models}.
\par
Specifically, instead of $\Omega \cos(\omega_0 t)$ as the transition term induced by the laser in the monochromatic case, we now have $(\Omega + {d W_{\Omega}}/{dt}) \cos(\omega_0 t + W_{\delta})$ when laser noise exists. $W_{\Omega} = s_{\Omega} W_1$ and $W_{\delta} = s_{\delta} W_2$ ($W_1$ and $W_2$ are independent Brownian motions) are scaled Brownian motions with scaling factors $s_{\Omega, \delta}$. They are independent of each other, implying that the noise in the intensity has nothing to do with the noise in the frequency. Brownian motion is not differentiable, hence the derivative notation $f = {d W_{\Omega}}/{dt}$ is interpreted in the sense of integration: $\int f dt = \int dW$. The stochastic calculus employed here is according to It\^o calculus \cite{Zoller}.
\par
The starting point is the Schr\"odinger equation for a two--level atom under the rotating--wave approximation, in the rotating--wave frame of the atomic resonance frequency. For ease of notation, let $\Omega' = \Omega + \frac{d W_{\Omega}}{dt}$, then
\begin{equation}
\label{schrodinger_eq}
i\frac{d}{dt} 
\begin{bmatrix}
c_1\\
c_2
\end{bmatrix}
=
\underbrace{
\frac{1}{2}
\begin{bmatrix}
0 & \Omega' e^{i W_{\delta}}\\
\Omega' e^{-i W_{\delta}} & 0
\end{bmatrix}
}_{\hat{H} = H/\hbar}
\begin{bmatrix}
c_1\\
c_2
\end{bmatrix},
\end{equation}
which allows the instantaneous energy eigenstate at time $t$, or dressed states:
\begin{equation}
\label{step_choice}
|+,t\rangle = 
\frac{1}{\sqrt{2}}
\begin{bmatrix}
1 \\
e^{-i W_{\delta}(t)}
\end{bmatrix}
\textrm{  ;  }
|-,t\rangle = 
\frac{1}{\sqrt{2}}
\begin{bmatrix}
1 \\
- e^{-i W_{\delta}(t)}
\end{bmatrix},
\end{equation}
which become stochastic and are different from the usual dressed states. The transform matrix between the dressed states and the bare states is given by:
\begin{equation}
B(t) = 
\frac{1}{\sqrt{2}}
\begin{bmatrix}
1 & e^{i W_{\delta}(t)} \\
1 & -e^{i W_{\delta}(t)}
\end{bmatrix}.
\end{equation}
\par
We now use the spirit of Feynman's path integral to analyze the time evolution associated with Eq.\eqref{schrodinger_eq}. SDE has close ties with Feynman's path integral, and this connection is further explored here for the case in which the basis states are discrete. We begin by constructing the propagator connecting the past $t_0$ to the future $t_n = T$. Let $\Theta = (1 - i \hat{H}(t_{n-1}) \Delta t ) \cdots  (1 - i \hat{H}(t_0) \Delta t)$ with $n$ time intervals $\Delta t = {(T - t_0)}/{n}$ between $t_0$ and $t_n$. Then the propagator is given by taking the limit of $\Delta t$ going to zero:
\begin{align}
\textrm{Pro}&\textrm{pagator} = \lim_{\Delta t \to 0 } \Theta \nonumber \\
&= \lim_{\Delta t \to 0 } (1 - i \hat{H}(t_{n-1}) \Delta t ) \cdots  (1 - i \hat{H}(t_0) \Delta t)
\label{propagator}
\end{align}
\par
To evaluate $\Theta$, we insert the identity operator $1 = |+, t_j\rangle \langle +, t_j| + |-, t_j\rangle \langle -, t_j|, j = 0,1,\cdots, n;$ into every adjacent time interval. Then $\Theta$ is transformed into a time--ordered product of a series of transition amplitudes: $g = g_{t_{n-1}} \cdots g_0$, where $\Theta = B^\dagger(T) g B(0)$ with respect to the bare states basis. Here $g_{t_j}$ is defined as
\begin{align}
g_{t_j} &= (|+, t_{j+1}\rangle \langle +, t_{j+1}| + |-, t_{j+1}\rangle \langle -, t_{j+1}| ) \nonumber\\
 &(1 - i \hat{H}(t_j) \Delta t)(|+, t_j\rangle \langle +, t_j| + |-, t_j\rangle \langle -, t_j| ).
\end{align}
\par
Here we regard $s_\delta$ as the ordering parameter and perform a perturbative calculation up to the first order in $s_\delta$. Then we apply It\^o's lemma to $e^{ - i W_\delta(\Delta t)} - 1$, keep the first order of $s_\delta$ and $\Delta t$  and arrive at Eq.\eqref{time_advance}. Therefore $g_{t_j}$ in the matrix representation is given in Eq.\eqref{g^t}.
\par
\begin{subequations}
\label{time_advance}
\begin{align}
        \langle +, t_{j+1} | +, t_j\rangle &\approx 1 + \frac{1}{2} i \Delta W_\delta;
        \langle +, t_{j+1} | -, t_j\rangle \approx -\frac{1}{2} i \Delta W_\delta \\
        \langle -, t_{j+1} | +, t_j\rangle &\approx -\frac{1}{2} i \Delta W_\delta;
        \langle -, t_{j+1} | -, t_j\rangle \approx 1 + \frac{1}{2} i \Delta W_\delta
\end{align}
\end{subequations}
\begin{equation}
\label{g^t}
g_{t_j} = 
\begin{bmatrix}
1 + \frac{1}{2} i \Delta W_\delta - \frac{1}{2}i \Omega' \Delta t & -\frac{1}{2} i \Delta W_\delta\\
-\frac{1}{2} i \Delta W_\delta & 1 + \frac{1}{2} i \Delta W_\delta + \frac{1}{2}i \Omega' \Delta t 
\end{bmatrix}
\end{equation}
\par
To proceed with the product of $g = g_{t_{n-1}} \cdots g_0$, we take the logarithm $\ln g = \ln(g_{t_{n-1}} \cdots g_0)$. For $\Delta t$ sufficiently small, $\ln g$ is as in Eq.\eqref{ln g_t} up to the first order.
\begin{equation}
\label{ln g_t}
\ln g_{t_j} = 
\begin{bmatrix}
\frac{1}{2} i \Delta W_\delta - \frac{1}{2}i \Omega' \Delta t & -\frac{1}{2} i \Delta W_\delta\\
-\frac{1}{2} i \Delta W_\delta & \frac{1}{2} i \Delta W_\delta + \frac{1}{2}i \Omega' \Delta t 
\end{bmatrix}.
\end{equation}
\par
If $\ln(g_{t_{j-1}} \cdots g_0) = \ln g_{t_{j-1}} + \cdots + \ln g_0$ holds, then in the process of letting $\Delta t \to 0$, we are essentially taking the integral of the building blocks of Eq.\eqref{ln g_t}. Suppose that $\lim_{\Delta t \to 0} g = G$ and the time starts at $t_0 = 0$, then we arrive at Eq.\eqref{G_time_evolution} in matrix format, and therefore we have $\textrm{Propagator} = B^\dagger(T) G B(0)$.
\begin{equation}
\label{G_time_evolution}
G = \exp{
\begin{bmatrix}
\frac{1}{2} i W_\delta - \frac{1}{2}i \Omega T - \frac{1}{2}i W_\Omega & -\frac{1}{2} i W_\delta\\
-\frac{1}{2} i W_\delta & \frac{1}{2} i W_\delta + \frac{1}{2}i \Omega T +  \frac{1}{2}i W_\Omega
\end{bmatrix}
}.
\end{equation}
\par
Before delving into the physical meaning of Eq.\eqref{G_time_evolution}, we check the validity of the condition $\ln(g_{t_{j-1}} \cdots g_0) = \ln g_{t_{j-1}} + \cdots + \ln g_0$. At first glance in Eq.\eqref{ln g_t}, it seems as if all $\ln g_{t_j}$ look the same and hence they should commute with each other. Yet, because of the Markovian property of Brownian motion, in fact $\Delta W_\delta$ at different times are independent, hence the commutativity argument fails.
\par
By the virtue of Baker--Campbell--Hausdorff formula, the expansion of $\ln(g_{t_{j-1}} \cdots g_0)$ contains infinite numbers of higher order terms of commutators besides $\ln g_{t_{j-1}} + \cdots + \ln g_0$. We verify that up to the first order that $\ln g_{t_{j-1}} + \cdots + \ln g_0$ equals $\ln(g_{t_{j-1}} \cdots g_0)$. This can be understood from two aspects. The first point is that, only terms in the first order of the scaling factor are kept, which is consistent with our starting point: a perturbative calculation up to the first order in the scaling factor, for both $W_\delta$ and $W_\Omega$. The second point is a bit more subtle. If we take the expectation values of $\ln g_{t_j}$, we see that all $E(\ln g_{t_j})$ commute with each other, where $E(\cdot)$ denotes taking the expectation; moreover, because of this independence, we have $E(g_{t_{j-1}} \cdots g_0) = E(g_{t_{j-1}}) \cdots E(g_0) $. In other words, the expectation value always evolves adiabatically in time, as expected.
\par
$G$ contains higher order terms of $s_\delta$ and $s_\Omega$, as can be seen from a Taylor expansion of Eq.\eqref{G_time_evolution}. However, according to our previous analysis $G$ is accurate to the first order and there is no guarantee that those higher order terms given in Eq.\eqref{G_time_evolution} are necessarily the right ones. Corrections to $G$ for the higher order terms can be made from higher order perturbation calculations.
\par
A few sanity checks for the derived propagator are shown in the following. From Eq.\eqref{G_time_evolution}  we see that $G^\dagger (T) G(T) = 1$, and hence the unitarity is preserved even under the presence of laser noise, which is expected. In the limit of the laser being noiseless ($s_\delta = 0$ and $s_\Omega = 0$), $G(T)$ reduces to 
$\exp
\begin{bmatrix}
-\frac{1}{2}i \Omega T & 0 \\
0 & \frac{1}{2}i \Omega T
\end{bmatrix}$,
 which is exactly the case of a two--level atom driven by a perfect on resonance laser.
\par
The spirit of the path integral is the summation over the contributions from all the paths to the transition amplitudes. Compared to the usual scheme of path integrals, the differences here are (1). At each time step the choices are discrete [Eq.\eqref{step_choice}]; (2). All the paths are stochastic. An intuitive picture is that the fluctuations of the laser drive all the paths to fluctuate, as a consequence the propagator as the summation over the contributions of all the paths fluctuates. Unless in very special cases, the fluctuations induced by laser noise in the paths do no cancel each other. Therefore in the end the detailed mechanism of how the laser fluctuates will determine stochastic property of the atomic wave function.
\par
The next step is to examine how laser noise changes the response of an atom to a $\pi$--pulse ($\Omega T = \pi$). If the interaction starts when the atom is in the ground state, the atomic wave function at time $T$ is
\begin{equation}
\begin{bmatrix}
c_1\\
c_2
\end{bmatrix}
=
B^\dagger(T) G(T) B(0) 
\begin{bmatrix}
1\\
0
\end{bmatrix},
\end{equation}
from which the population in the excited states $c^*_2 c_2$ can be computed as
$\begin{bmatrix}
1 & 0
\end{bmatrix}
B^\dagger (0) G^\dagger (T) B(T)
\begin{bmatrix}
0 & 0\\
0 & 1
\end{bmatrix}
B^\dagger (T) G (T) B(0)
\begin{bmatrix}
1 \\ 0
\end{bmatrix}$.
The result at time $T$ is: 
\begin{equation}
\label{pi_pulse_1}
c^*_2 c_2 = n^2_x \sin^2(\phi / 2),
\end{equation}
where $n_x$ is the $x$--component of the vector $\vec{n} = -((\Omega T + W_\Omega (T)),0,W_\delta (T))/\phi$ and $\phi = \sqrt{W^2_\delta (T) + (\Omega T + W_\Omega (T))^2}$. In the limit of a noiseless laser ($s_\delta = 0$ and $s_\Omega = 0$) Eq.\eqref{pi_pulse_1} reduces to the usual Rabi oscillation $\sin^2(\Omega T/2)$. However, because of the existence of laser noise, at the end of a $\pi$--pulse the population in the excited state is now a random variable rather than a deterministic value $1$. By measuring its statistical properties, the information about the laser noise can be revealed, as is implied by Eq.\eqref{pi_pulse_1}. For example, if the intensity fluctuation dominates the laser noise, then after a $\pi$--pulse $c^*_2 c_2$ is approximately $\sin^2((\pi + W_\Omega (T))/2) = 1/2 + 1/2 \cos(W_\Omega (T))$ where the laser intensity fluctuation will be adequately described by repeatedly recording $c^*_2 c_2$.
\par
The form of the random variable $\sin^2((\pi + W_\Omega (T))/2)$  is because of our choice of the intensity fluctuation model, in which the integral of the Rabi frequency over time is a Brownian motion plus a constant drift. If we model the intensity fluctuation by an Ornstein--Uhlenbeck process $X(t)$, then after a $\pi$--pulse $c^*_2 c_2$ is $\sin^2(\int_0^T \frac{1}{2}X(t)dt)$, where $\int_0^T X(t)dt$ obeys a distribution with mean $\pi$ and variance corresponding to the intensity noise strength. This is consistent with our motivation: we are not interested in modeling different kinds of laser noise; rather, we provide a framework showing how laser noise would manifest itself in the laser--atom interaction, and how the laser noise's stochastic properties get written into the atom wave function during time evolution.
\par
The laser noises are typically regarded as Markovian, and henceforth no information about the history prior to time $t_0$ can be extracted from a measurement performed after $t_0$. Then we expect that the stochastic properties of of the atom wave function within a time interval solely depend on the laser noise's behavior during that time interval. Therefore, the time correlation of the atom wave function provides an insight into the stochastic nature of the laser noise during a certain time period. Many types of correlations can be constructed for different purposes. Here for an example we are looking into the inner product of the wave functions at different times $T_1$ and $T_2$,  $c^*_2(T_2) c_2(T_1) +c^*_1(T_2) c_1(T_1)$, which can be explicitly evaluated as
$\begin{bmatrix}
1 & 0
\end{bmatrix}
B^\dagger(0) G^\dagger (T_2) B(T_2)
B^\dagger(T_1) G(T_1) B(0)
\begin{bmatrix}
1 \\ 0
\end{bmatrix}
$. 
The result up to the first order in $s_\delta$ and $s_\Omega$ is
\begin{equation}
\label{correlation_1}
e^{- \frac{i}{2} W_{\delta}(\Delta T)} (\cos (\frac{\phi}{2}) - i n_z \sin (\frac{\phi}{2})),
\end{equation}
where $\Delta T = T_2 - T_1$, $n_z = -\frac{W_\delta(\Delta T)}{\phi}$, and $\phi = \sqrt{W^2_\delta(\Delta T) + (\Omega \Delta T + W_\Omega (\Delta T))^2}$. In the limit of the laser being noiseless ($s_\delta = 0$ and $s_\Omega = 0$) Eq.\eqref{correlation_1} reduces to $\cos (\frac{1}{2} \Omega \Delta T)$, which is exactly the case of an atom driven by a perfect sinusoidal wave. If, for example, the phase noise (described by the phase diffusion model) dominates and the Rabi oscillation is fast compared to the dephasing rate ($\Omega \Delta T \gg \| W_\delta (\Delta T) \|$), then Eq.\eqref{correlation_1} is approximately $(1 - \frac{i}{2} W_\delta(\Delta T)) \cos(\frac{1}{2}\Omega \Delta T)$. This simplified form leads to several consequences. The correlation defined as the expectation value $E((1 - \frac{i}{2} W_\delta(\Delta T)) \cos(\frac{1}{2}\Omega \Delta T))$ is $\cos(\frac{1}{2}\Omega \Delta T)$, which again coincides with the case of no laser noise at all, though it does have a non--zero variance. When $\Omega \Delta T = (2N+1) \pi, N = 0, 1, 2, \cdots$, this inner product of wave functions is immune to the laser phase noise to the first order. When $\Omega \Delta T = 2N \pi, N = 0, 1, 2, \cdots$, the fluctuation $\frac{i}{2} W_\delta(\Delta T)$ in this inner product of wave functions is proportional the phase diffusion of the laser noise itself.
\par
Finally we want to discuss how the previous discussions connect to the notion of laser linewidth and related experimental measurements. Laser linewidth is a repeatedly discussed topic in literature. Good theoretical references can be found at \cite{PhysRev.112.1940} and \cite{PhysRev.161.350}. E. D. Hinkley and Charles Freed \cite{PhysRevLett.23.277} presented the early experimental efforts to measure the linewidth and a thorough understanding of the nature of this kind of measurement. Here we plan to use probability method as a tool to show that the simple and direct model of laser noise can lead to experimentally observed laser linewidth. In particular, we are going the describe the line shape in a heterodyne experiment that beats two lasers with frequency noises described by the phase diffusion model.
\par
The intensity of the beat of two lasers of the same type with frequency noises described by the phase diffusion model  $\cos(\omega_1 t + s W_1(t))$ and $\cos(\omega_2 t + s W_2(t))$ is given in Eq.\eqref{beat_1}.
\begin{align}
& (\cos [ \omega_1 t + W_1(t) ] + \cos [\omega_2 t + W_2(t) ])^2 \nonumber \\
&= 2 + \cos[ (\omega_1 - \omega_2)t + (W_1 - W_2)] \nonumber \\
&+ \cos[ (\omega_1 + \omega_2)t + (W_1 + W_2)],
\label{beat_1}
\end{align}
where $W_1$ and $W_2$ are independent Brownian motions, and $s$ is the scaling factor. A detector would respond to the term at the beat frequency $\delta = \omega_1 - \omega_2$ in Eq.\eqref{beat_1}. We rewrite that term as $\cos(\delta t + s_0W_0)$, where $s_0W_0 = s(W_1 - W_2)$, $W_0$ is a new Brownian motion and $s_0 = \sqrt{2} s$. Analyzing the spectrum of the detector response is essentially calculating the averaged Fourier spectrum of $\cos(\delta t + s_0W_0)$, which is $E(\int_0^T \exp(i \omega t ) \cos(\delta t + s_0W_0) dt)$.
\par
Let $f(W_0(t), t) = \exp(i \omega t) \exp (i s_0W_0(t))$ and then apply It\^o's lemma to $f$:
\begin{align}
e^{i\omega T} & e^{i s_0 W_0(T)} = 1 + \int_0^T i s_0 e^{i\omega t} e^{i s_0 W_0(t)} W_0(dt) + \nonumber \\
&\int_0^T \{ -\frac{1}{2} s_0^2 e^{i\omega t} e^{ i s_0 W_0(t)} + i \omega e^{i\omega t} e^{ i s_0 W_0(t)} \} dt
\label{Ito_lemma_1}
\end{align}
Take the expectations on both sides of Eq.\eqref{Ito_lemma_1}, and note that the expectation of an It\^o integral is zero. Then
\begin{equation}
E(e^{i\omega T} e^{i s_0 W_0(T)}) = 1 + 0 + (i\omega - \frac{1}{2} s_0^2) E(\int_0^T e^{i\omega t} e^{ i s_0 W_0(t)} dt).
\end{equation}
$W_0(T)$ is a Gaussian random variable with variance $T$. Hence $E(e^{i s_0 W_0(T)}) = e^{-\frac{s_0^2 T}{2}}$. Then
\begin{equation}
\label{lineshape 1}
 E(\int_0^T e^{i\omega t} e^{ i s_0 W_0(t)} dt) = \frac{e^{i\omega T} e^{-\frac{s_0^2 T}{2}} - 1}{i\omega - \frac{1}{2} s_0^2}.
\end{equation}
The long expected result can now be calculated as
\begin{align}
E&(\int_0^T \exp(i \omega t ) \cos(\delta t + s_0 W_0) dt) = \nonumber \\
& \frac{1}{2}\{ \frac{e^{i(\omega + \delta) T} e^{-\frac{s_0^2 T}{2}} - 1}{i(\omega + \delta) - \frac{1}{2} s_0^2} 
+ \frac{e^{i(\omega - \delta) T} e^{-\frac{s_0^2 T}{2}} - 1}{i(\omega - \delta) - \frac{1}{2} s_0^2} \}.
\label{lineshape 2}
\end{align}
Eq.\eqref{lineshape 2} clearly shows a double-peak structure with some line shape. Let us just look at one branch
\begin{equation}
\frac{e^{i(\omega + \delta) T} e^{-\frac{s_0^2 T}{2}} - 1}{i(\omega + \delta) - \frac{1}{2} s_0^2} = \frac{(e^{i(\omega + \delta) T} e^{-\frac{s_0^2 T}{2}} - 1) (\frac{1}{2} s_0^2 + i(\omega + \delta))}{(\omega + \delta)^2 + (\frac{1}{2} s_0^2)^2}.
\label{lineshape 2.5}
\end{equation}
A technical point about such an experiment is that the device is usually performing the discrete version of the Fourier transform. As a result, for all the discrete $\omega$ values in the outcome, $\omega T$ is always a multiple of $2\pi$. Therefore, $e^{i\omega T}$ is always $1$. Taking the absolute value of Eq.\eqref{lineshape 2.5}
\begin{equation}
\label{lineshape 3}
|\frac{e^{i(\omega + \delta) T} e^{-\frac{s_0^2 T}{2}} - 1}{i(\omega + \delta) - \frac{1}{2} s_0^2} | = | e^{i\delta T} e^{-\frac{s_0^2 T}{2}} - 1 | \frac{1}{\sqrt{{(\omega + \delta)^2 + (\frac{1}{2} s_0^2)^2}}}.
\end{equation}
\par
Eq.\eqref{lineshape 3} explicitly shows: the intensity of the Fourier transform spectrum of the beat signal has the line shape of a Lorentzian. The linewidth is then $\frac{1}{2} s_0^2$, and this example shows the establishment of the relation between the parameters of the laser noise model in describing its effect on laser--atom interaction and the outcome of a typical laser linewidth measurement.
\par
All together, Eq.\eqref{G_time_evolution}, Eq.\eqref{lineshape 2} and Eq.\eqref{lineshape 3} have established the correspondence between the effects caused by the laser noise in the laser--atom interaction and the laser beating experiment, where the laser noise is described by the same model and parameter $s_0$. This correspondence can be further interpreted that we are comparing two frequencies in both situations: A noisy laser is compared with another noisy laser in the beat experiment of Eq.\eqref{beat_1}, while a noisy laser is compared to a noiseless atom in the laser--atom interaction of Eq.\eqref{schrodinger_eq}.
\par
In summary, we have shown by probabilistic methods the effects caused by laser noise in an infinitely narrow atomic transition. The randomness inherent in the laser noise would be transcribed to the atomic wavefuction which is then made into a random variable, whose behavior is determined by the nature of the laser noise.

\begin{acknowledgments}
We thank our advisors Harold Metcalf and Chris H. Greene for valuable discussions. 
This work has been supported in part by ONR and NSF. 
\end{acknowledgments}



\bibliographystyle{unsrtnat}

\renewcommand{\baselinestretch}{1}
\normalsize

\bibliography{ysref}
\end{document}